\documentclass[12pt]{article}
\usepackage{cmb,epsfig}
\newcommand{\degr}{^\circ}
\begin{document}
\heading{RADIO FREQUENCY FOREGROUND EMISSION\\ AND THE SEPARATION OF MICROWAVE\\
ANISOTROPIES FROM FOREGROUND CONTAMINATION}

\author{A.N. Lasenby} {MRAO, Cavendish Laboratory} { Cambridge, U.K.}

\begin{abstract}{\baselineskip 0.4cm 
The effects on CMB measurements 
of foreground contamination due to synchrotron radiation, free-free
emission and discrete sources are considered. Estimates of the level and power
spectrum of the Galactic fluctuations are made using low frequency maps and
recent Tenerife data. New methods for achieving a frequency-based separation of
Galactic and CMB components are discussed, and a positive/negative maximum
entropy algorithm shown to be particularly powerful in this respect.}

\end{abstract}

\section{Introduction}

A common feature of all current and future CMB experiments is that
contamination of the CMB signal by foreground components must be dealt with.
The most important contaminants over the frequency range of interest for the CMB
are the effects of discrete radio and sub-mm sources, and the effects of
our own Galaxy. Discrete sources can in principle be discriminated against on
the basis of their quite different angular size compared to the CMB fluctuations of
interest. For Galactic effects however, no such assumption can be made, and we
must discriminate between them and CMB purely on the grounds of their differing
behaviour with frequency.

In this context it is very important to have as good an estimate as we can
of the likely level of Galactic anisotropies, and to be armed with good
algorithms for separating out their effects. This is an area which will see
great development over the next few years as we prepare for the next
generation of experiments, but as a preliminary effort in this direction we
consider here, in schematic form, some recent estimates of Galactic
variation gathered using current CMB experiments and low frequency maps. Data
on a new type of discrete source whose spectrum can mimic a thermal source is
also discussed. In the main part of the contribution, methods of performing the
Galactic frequency separation are briefly reviewed, and the good performance
for this purpose of a maximum entropy algorithm allowing both positive and negative
fluctuations is highlighted.

\section{Power spectrum and level of Galactic fluctuations}

It is difficult to use the low frequency 408 MHz \cite{haslam} and 1420 MHz
\cite{reich} maps to extrapolate Galactic variations to CMB observing
frequencies. The problems include difficulties with baselevels, scanning
effects and of course with only two frequencies available only one spectral
index can be estimated, whereas in principle the maps are mixtures of
synchrotron and free-free emission, which have different spectral indices. As
further problems, the synchrotron spectral index probably varies with frequency
(on physical grounds it is expected to steepen with increasing frequency), and
the maps are of a limited resolution ($0.85\degr$ and $0.6\degr$ at 408 and
1420 MHz respectively) which makes it difficult to assess effects on scales
below where the first Doppler peak is expected to lie for CMB fluctuations in
an inflationary picture.

\begin{figure}[t]
\centerline{\hbox{\epsfig{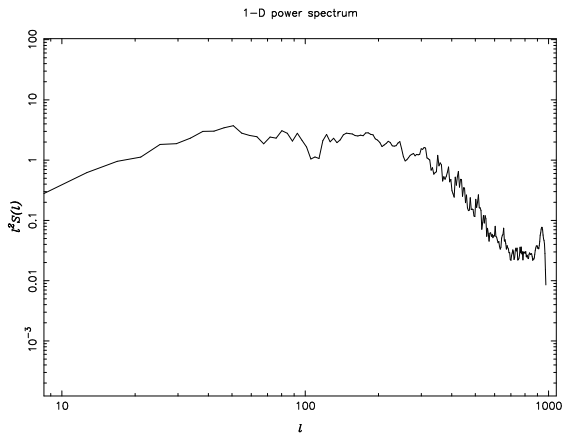}}}
%\vspace*{2.9in}
\caption{1-d power 408 MHz spectrum for Tenerife patch}
\label{fig:tenerife_408}
\end{figure}

\begin{figure}[t]
\centerline{\hbox{\epsfig{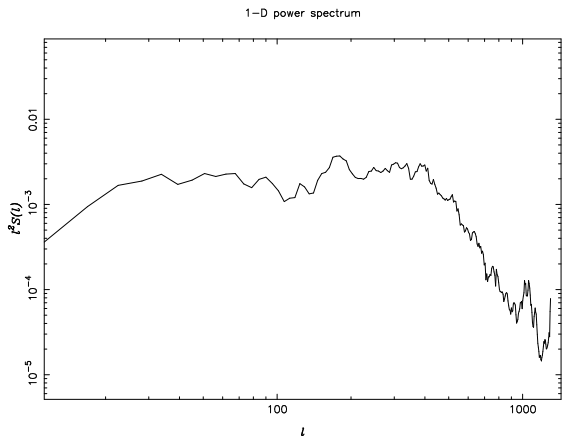}}}
%\vspace*{2.9in}
\caption{1-d power 1420 MHz spectrum for Tenerife patch}
\label{fig:tenerife_1420}
\end{figure}

However, one thing we {\em can\/} use the low frequency maps for is to estimate
the spatial power spectrum of high-latitude Galactic variations, at least on
angular scales above the point where scanning effects and/or the limited
resolution become significant. Figs.~\ref{fig:tenerife_408} and
\ref{fig:tenerife_1420} (prepared by M.P. Hobson) show the 1-dimensional
spatial power spectra at 408 and
1420 MHz for a patch of sky which has been extensively sampled by the
`Tenerife' experiments. The section used is $165\degr$ to $225\degr$ in Right
Ascension and $35\degr$ to $45\degr$ in Declination. A Hanning window has been
applied and an overall mean subtracted. The spectrum is given in terms of
spherical multipole $l$ and multiplied by $l^2$, so that it can be directly
compared to the usual plots of CMB power --- in particular, a scale invariant
spectrum would have zero slope on this plot. It has recently been suggested, on
a number of grounds, that the Galactic power spectrum behaves as $l^{-3}$ as
one approaches smaller angular scales \cite{tegmark}.  The observational basis
of this for dust emission seems well established, but the plots presented here
favour a slightly less steep index for synchrotron emission (assuming this is
the dominant component at 408 and 1420 MHz). In fact, the fall off at higher
$l$ away from an $l^{-2}$ scale-invariant behaviour, begins in both plots at an
angular scale which corresponds to the effects one would expect given the
limited resolution of the maps. In any case, these results make it clear that the
Galactic foreground problem will not get any {\em worse\/} compared to the CMB
levels as one goes to higher resolutions, and this is encouraging for future
experiments.

\subsection{Estimating the fluctuation level}
The Tenerife experiments constitute a suite of three instruments, operating at
10, 15 and 33 GHz, on angular scales of $\sim 5\degr$, and are therefore potentially
well-suited to providing an estimate of the Galactic component on these scales.
They were designed and built at Jodrell Bank \cite{davies92,dec40-I} and are
operated by the IAC in-situ on Tenerife island. In fact, the level of
anisotropy seen does not change very much in passing from 15 to 33 GHz \cite{hannat94}
and this makes it difficult to assess the Galactic contribution, the
implication being that the majority of the signal detected is CMB. However, we
can set upper limits on the Galactic contribution as follows. Using the COBE
DMR 2 year data and an assumed scale invariant CMB spectrum, we can predict a
level of $31 {\rm \, \mu K}$ that should be the {\em rms} observed in the Tenerife
scans (at all frequencies since the beam geometry is the same at each
frequency). We can then estimate other contributions to the signal (in
particular the point source contribution) and then subtract in quadrature from
the {\em observed\/} {\em rms\/} in order to obtain a residual, which can be
interpreted as an estimate of the Galactic contribution. This process is shown
in Table~\ref{ta:results}, with the result and 68\% error bounds given in the
\begin{table}[h]
\begin{center}
\begin{tabular}{|c|c|c|}
\hline \hspace*{.2in} Component \hspace*{.2in} & 10GHz signal & 15GHz signal \\ 
     &  $\mu$K      & $\mu$K    \\ \hline
Point sources & 12.8 & 6.5 \\ 
Signal & $39^{+16}_{-14}$ & $37^{+11}_{-8}$ \\
CMB pred. ($n=1$) & 31 & 31 \\
Residual & $24^{+22}_{-24}$ & $20^{+16}_{-20}$ \\
1420 MHz $\beta=2.75$ & 89 & 33 \\
408/1420 MHz & 185 & 76 \\
$\beta$ from 1420 MHz & 3.4 & 3.0 \\
\hline
\end{tabular}
\end{center}
\caption{Breakdown of different contributions to the measured anisotropy in the
Tenerife experiments at 10 and 15 GHz.}
\label{ta:results}
\end{table}
row labelled `residual'. In the next two rows, the predicted Galactic contribution
is shown assuming first a spectral index ($T\propto \nu^{-\beta}$) of
$\beta=2.75$ in going from 1420 MHz to the Tenerife frequencies and secondly using
the 408 and 1420 MHz maps themselves to provide the spectral index to use in
the extrapolation. In the final row, the $\beta$ necessary to get the residuals
found starting from 1420 MHz is shown, and indicates a considerable steepening
over the canonical 2.75 value. The area of sky included in generating these
figures is the same as used for the power spectral plots shown above. 

The fact that the residuals do not change too much between 10 and 15 GHz
suggests that they do not represent a true Galactic level. (For example, even
wholly free-free would reduce by a factor $\sim 2.3$ in passing from 10 to 15
GHz.) However, we could takes these numbers as `worst case' levels. Thus, if we
assume a free-free spectrum and take the numbers at 15 GHz, we can now predict
the levels that will be seen in the COBRAS/SAMBA lower channels on the $\sim
5\degr$ scale. 
\medskip
\begin{center}
\begin{tabular}{cccc}
$\nu$(GHz) & 31.5 & 53 & 90 \\ 
$\Delta T_{\rm Gal} ({\rm \, \mu K})$ & $< 4^{+4}_{-4}$ & $<1.4^{+1.1}_{-1.4}$ & $<
0.46^{+0.37}_{-0.46}$ \\ 
\end{tabular}
\end{center}
\medskip
We can compare these levels with the ($|b|>20\degr$) all-sky 
free-free levels suggested by Kogut's analysis (this
volume) of the COBE DMR -- DIRBE 140 $\mu$m cross-correlation:
\medskip
\begin{center}
\begin{tabular}{cccc}
$\nu$(GHz) & 31.5 & 53 & 90 \\ 
$\Delta T_{\rm ff} ({\rm \, \mu K})$ & $21\pm 5$ & $7\pm 2$ & $2.3\pm 0.5$ \\ 
\end{tabular}
\end{center}
\medskip
This suggests that the Tenerife region (approx 600 square degrees) is $\sim 5$
times better than average over the 
$|b|>20\degr$ sky. The availability of such large patches with relatively low
Galactic levels is encouraging for future ground-based and balloon experiments,
especially when coupled with the conclusion that things get no worse as one
moves down in angular scale from the Tenerife/COBE scales.

\section{Discrete sources}

\begin{figure}[t]
\centerline{\hbox{\epsfig{figure=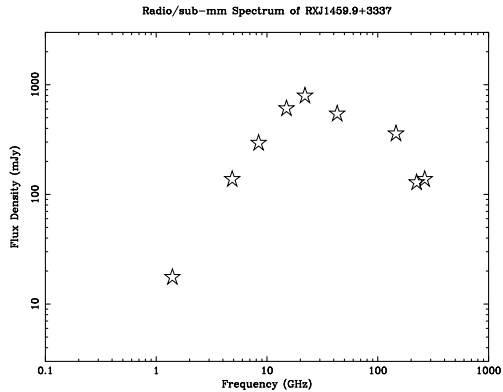,height=2.9in,angle=-90}}}
%\vspace*{2.9in}
\caption{The radio/sub-mm spectrum for the X-ray selected quasar
RXJ1459.9+3337. (Figure courtesy of A. Edge)}
\label{fig:gips}
\end{figure}

As mentioned above, discrete radio and sub-mm sources can be discriminated from
CMB fluctuations via their quite different angular scale. This requires having higher
resolution available than the resolution at which the CMB information is
desired of course. As an example of how trying to eliminate radio sources
using spectral information alone could go wrong, we show in Fig.~\ref{fig:gips} the
spectrum of the source RXJ1459.9+3337 over the radio sub-mm region. (This
figure and associated information were kindly provided by Alistair Edge.)
The data are from multi-frequency VLA
observations taken in April 1996 and JCMT photometry in
May 1996. RXJ1459.9+333 is a quasar that was discovered first in X-rays, and
only later measured as a radio source. The spectrum `turns over' at a point
somewhere between 20 and 40GHz, and on the rising side has a slope of $\sim
1.8$, virtually indistinguishable from the CMB over this range. Since the source
was selected at X-ray, rather than on the basis of its radio properties, we do
not how how rare such sources are --- it would not have turned up in previous
radio surveys. Some indication of the frequency of occurrence of such sources
may come from another such source, slightly weaker, which was found serendipitously in a
search of $\sim 0.5 $ square degrees in a ROSAT cluster field using the Ryle
Telescope at 15GHz. One such source in each $\sim 0.5 $ square degrees area
would not pose a formidable threat to CMB astronomy, but shows that we do have
to be on our guard for the unexpected.

\section{Algorithms for frequency separation}

Methods for separating foreground contamination from the CMB include
pixel by pixel separation using least squares \cite{brandt},
marginalization \cite{hobson,scott}, Wiener filtering \cite{bouchet,tegmark}
and maximum entropy methods (MEM) \cite{jones,maisinger}. Here we wish
to consider particularly the use of 2-channel (positive/negative)
maximum entropy, which can simultaneously achieve both deconvolution and
frequency separation.

The classical MEM approach to combining observations, and removing the
effects of a switched beam etc., uses an assumption of positivity (e.g.
White and Bunn \cite{whitebunn}). The CMB sky, however, apart from an
unobservable monopole, has zero mean and is expected to fluctuate
equally between positive and negative. White and Bunn try to repair this
by adding a positive constant to the data, but the need to include a
default measure as well results in less space in which to reconstruct
negative as opposed to positive features, and a consequent bias in the
results. This may be verified by numerical simulations. Instead 2-
channel MEM may be used, in which positive and negative channels are
reconstructed simultaneously. This has been shown to work well in
deconvolution of switched beam data \cite{jones} and avoids the
problem with bias. Recently, Maisinger {\em et al.\/} \cite{maisinger}
have applied this
technique to foreground/CMB separation using multifrequency data in the
context of simulated interferometer observations, and the results
below  are taken from that paper. This method can be contrasted with
that of Wiener filtering. The latter provides the optimal linear method
in the case that the power spectrum of the separate components is known. 
MEM is a non-linear method, which does not rely on any assumption about
the power spectra, but just that the skies it reconstructs have the
smallest information content consistent with the actual data.

\begin{figure}[t]
\centerline{\hbox{\epsfig{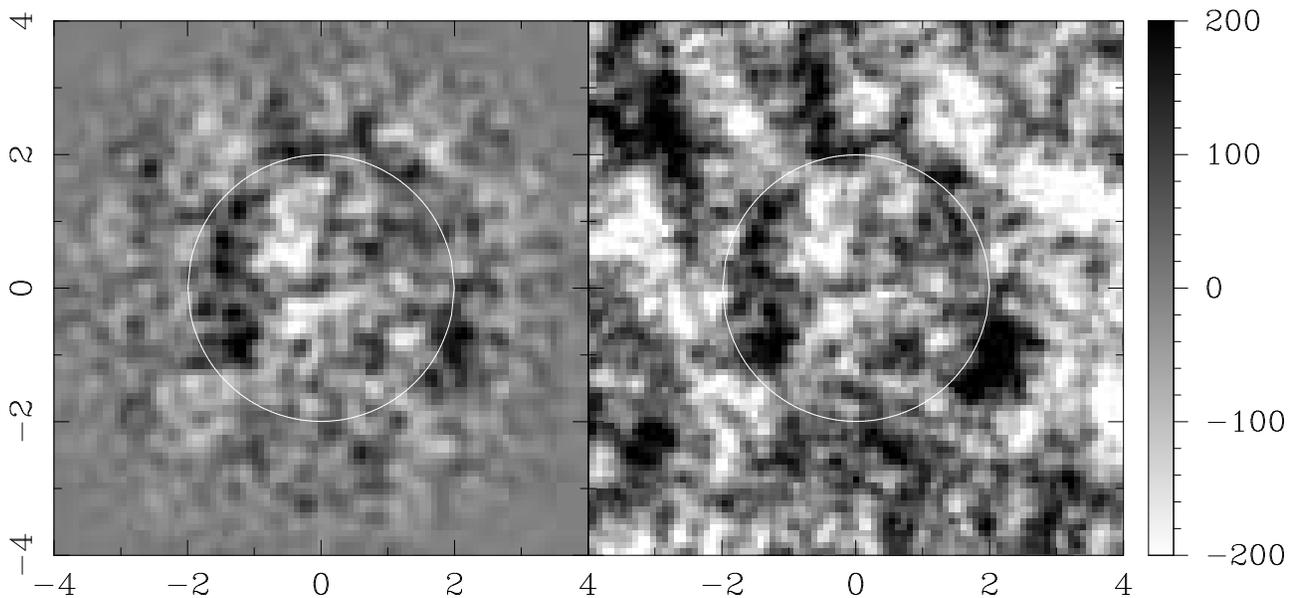}}}
%\vspace*{2.9in}
\caption{The MEM reconstruction of the CMBR component at 30~GHz in a
$8\degr\times8\degr$-field compared to the true underlying CMBR map
used in the simulation.  The grey-scales denote temperature
fluctuations in $\mu$K.  Neither of the maps is convolved with a
synthesised beam, and the image is kept on the reconstruction
grid. The circles indicate the FWHM of the primary telescope beam.}
\label{fig:cmbhigh}
\end{figure}

\begin{figure}[t]
\centerline{\hbox{\epsfig{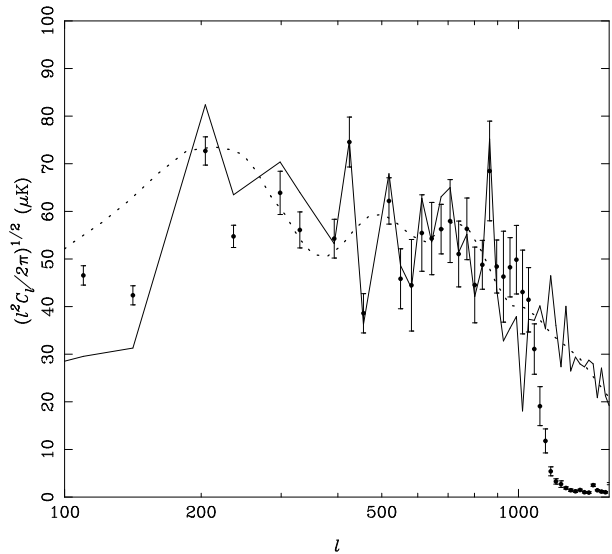}}}
%\vspace*{2.9in}
\caption{Comparison of power spectra of original (solid line) 
and reconstructed (points) CMBR maps. The error-bars indicate the
68-percent confidence limits as derived from 100 Monte-Carlo
simulations with different noise realisations. Also plotted is the
ensemble average CDM power spectrum (dotted line)
from which the input map was generated.}
\label{fig:cmbpower}
\end{figure}

\begin{figure}[t]
\centerline{\hbox{\epsfig{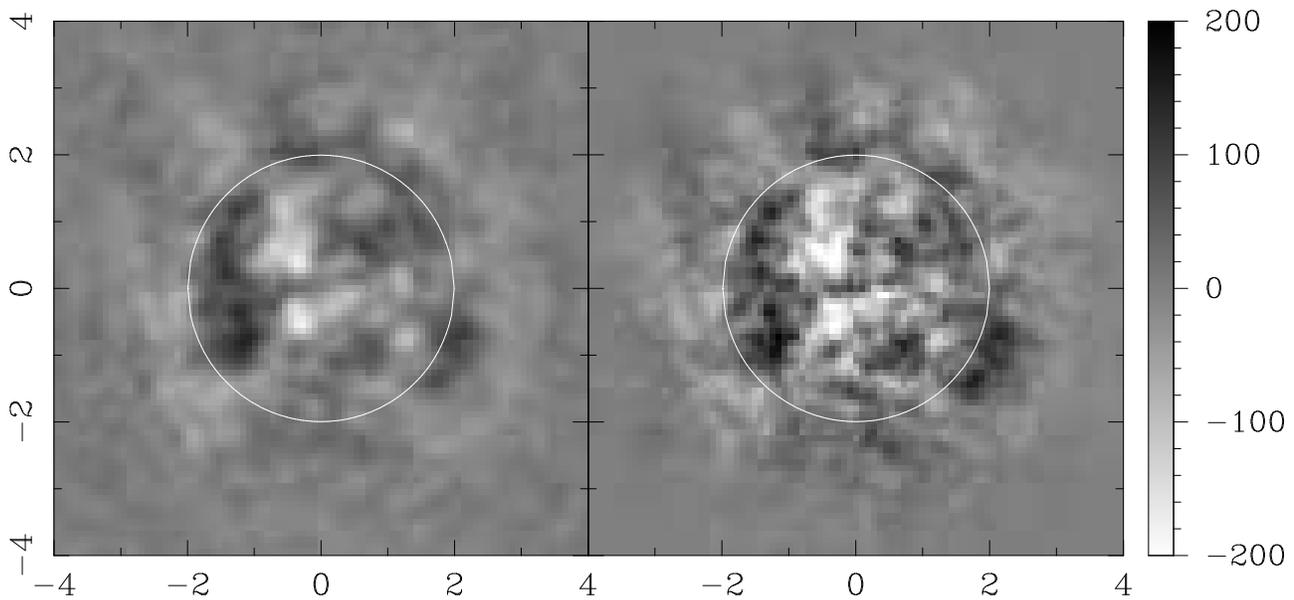}}}
%\vspace*{2.9in}
\caption{The Wiener filter reconstruction of the CMBR component at
30~GHz in a $8\degr\times8\degr$-field compared to the true underlying
CMBR map used in the simulation, multiplied by the primary beam. 
In contrast to Fig.~\ref{fig:cmbhigh}, no deconvolution has been
attempted in this case. The circles indicate the FWHM of the primary
telescope beam.}
\label{fig:wiencmb}
\end{figure}

\begin{figure}[t]
\centerline{\hbox{\epsfig{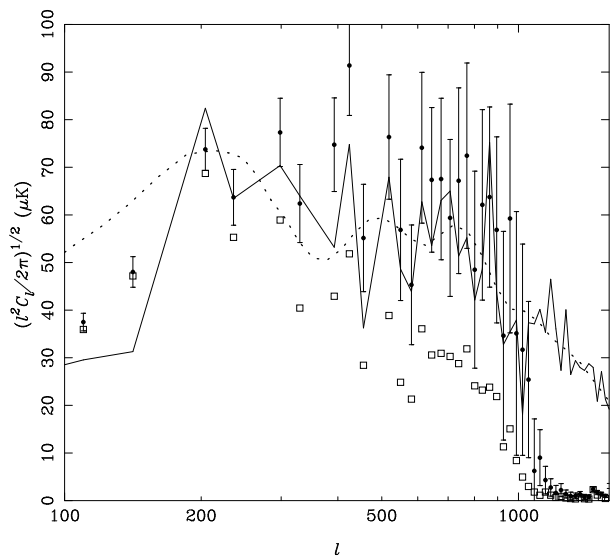}}}
%\vspace*{2.9in}
\caption{Comparison of power spectra of original (solid line) and
reconstructed CMBR maps as obtained from linear filtering methods.
Power spectra for both Wiener filtering (squares) and the rescaling
method (dots) are plotted.  Also plotted is the ensemble average
CDM power spectrum (dotted line) from which the input map was
generated.}
\label{fig:wiencmbpower}
\end{figure}

The examples given in the following figures concentrate on the case
of the VSA (see Jones, this volume). Two channels of data are used, at
28 and 38 GHz, each with bandwidth 2 GHz, with observations made for 
$ 30 \times $ 12 hours at each frequency. Fig.~\ref{fig:cmbhigh}
shows the true (CDM) sky used in the simulation on the right, and its
MEM reconstruction on the left. An important point which must be made is
that the reconstruction can only be made within the sky area sampled by
the telescope, and this is controlled by the circular primary beam. This
is why the reconstruction fades to zero (the default) near the edges.
In Fig.~\ref{fig:cmbpower} the power spectrum recovered from the
reconstructed map is shown together with the real power spectrum and
errors estimated by Monte Carlo simulations. The real power spectrum of
course deviates from the underlying ensemble CMB power spectrum since
only the realization corresponding to the single sky patch used is
available. Figs.~\ref{fig:wiencmb} and \ref{fig:wiencmbpower} show the
same but for an implementation of the Wiener filtering approach to this
problem. Note in each case (MEM and Wiener), the true Galactic frequency
spectral index used in the simulations ($\beta=2.7$) is assumed known by the
algorithm. However, it is not too difficult to relax this assumption, for
example by including {\em two\/}  channels for frequency reconstruction,
each with a different assumed spectral index, and letting the algorithms decide
how much weight to give each channel. 

Overall it can be seen that this application of MEM is very successful,
clearly outperforming the Wiener method in this case.
We are investigating its use for frequency based foreground/CMB
separation in a variety of other contexts also.


\acknowledgements{I am grateful to all the members of the CAT and Tenerife teams 
at Jodrell Bank and Cambridge for help with the preparation of material, particularly
Mike Hobson, Stephen Hancock, Mike Jones, Klaus Maisinger, Carlos Guti\'errez 
and Aled Jones.
}

\vfill
\end{document}